\documentclass[apsspec,showpacs,twocolumn,nofootinbib,groupedaddress,lengthcheck,preprintnumbers]{revtex4-1}
\usepackage{amsmath}
\usepackage{soul}
\usepackage{amssymb}
\usepackage{graphics}
\usepackage{slashed}
\usepackage[dvips]{graphicx}
\usepackage{graphicx}
\usepackage[usenames]{color}

\def\simge{\mathrel{%
       \rlap{\raise 0.511ex \hbox{$>$}}{\lower 0.511ex \hbox{$\sim$}}}}
\def\simle{\mathrel{
       \rlap{\raise 0.511ex \hbox{$<$}}{\lower 0.511ex \hbox{$\sim$}}}}

\newcommand \beq{\begin{eqnarray}}
\newcommand \eeq{\end{eqnarray}}

\usepackage[colorlinks=true,linktocpage=true,linkcolor=blue,citecolor=blue]{hyperref}

\begin{document}

\title{Resolution of the hyperfine puzzle and its significance for two-fermion Dirac atoms}

\author{Gordon Baym$^{a}$ and Glennys R. Farrar$^c$}
\affiliation{
\mbox{$^a$Illinois Center for Advanced Studies of the Universe}\\
and\\
\mbox{Department of Physics, University of Illinois, 1110
  W. Green Street, Urbana, IL 61801} \\ 
\mbox{$^b$Center for Cosmology and Particle Physics, Department of Physics, New York University, NY, NY 10003}\\
}

\date{\today}

\begin{abstract}

    If we could shrink the ground state wave function of a hydrogen atom to a size $R$, at fixed electron mass and charge,  the 
 singlet hyperfine attraction in a spin singlet state would scale as $1/R^3$ -- faster than the increase in kinetic energy -- raising the question of why the hyperfine interaction does not lead to collapse of hydrogen, or similarly, positronium.   We resolve this issue within the framework of the full Dirac equation for the hydrogen atom ground state, employing 
a minimax variational calculation.  The result is that in a variational state of size $R$,  the magnetic moment of the electron assumes its usual value, $e\hbar/2mc$, when $R$ minimizes the total energy.   However, for $R<\hbar/mc$ there is no energy minimum, and the effective electron magnetic moment becomes essentially $eR/2$, softening the hyperfine interaction compared with the zero point energy.  Extending the Dirac variational calculation to Coulombic atoms with two oppositely charge pointlike fermions of arbitrary masses, e.g., positronium or muonium, the effective magnetic moment of the positive particle is similarly suppressed in small atomic configurations, leading to stability against collapse, while in hydrogen, the finite size of the proton is enough to ensure stability.   This same framework is, in addition, applicable in treating diquarks as relativistic Coulombic systems in the presence of color electric and magnetic interactions. 
  \end{abstract}

\maketitle

\section{Introduction}

 The hyperfine interaction \cite{fermi,breit} is cental to high-precision atomic spectroscopy \cite{horbatsch}, as well as in positronium \cite{KK} and the color magnetic interactions of quarks in hadrons \cite{drgg,mitbag}.   In a singlet spin state, the hyperfine interaction is attractive, and as Fermi showed, nominally proportional to the square of the wave function at the origin, giving a contribution to the energy  $\propto -1/R^3$, where $R$ is the size of the bound state.  Why does this interaction not lead to collapse of the bound state to $R\to 0$, since the kinetic energy rises only as $1/R^2$ non-relativistically  (and as $1/R$ relativistically) in the small $R$ limit.  Why do hydrogen and positronium not collapse?   
          
   The simplest picture of non-relativistic hydrogen illustrates the problem.  We assume a trial ground state wave function, 
  \beq
  \varphi(r) = \frac{e^{-r/R}}{\sqrt{\pi R^3}}; 
  \label{phicoul}
\eeq
which is the exact solution for the normal hydrogen ground state when the variational parameter $R$ equals the Bohr radius, $a_0 = \hbar/me^2$, with $m$ the electron mass.   In this state the expectation value of the kinetic energy is $1/2mR^2$, and the expectation value of the Coulomb energy is $-\alpha/R$, where $\alpha = e^2/\hbar c$ is the QED fine structure constant.  (We work in units $\hbar = c = 1$.)  In the absence of the hyperfine interaction, the expectation value of the total energy, 
\beq
  E_0(R) = \frac{1}{2mR^2} - \frac{\alpha}{R},
  \label{E0R}
\eeq
has the expected minimum, $-\alpha/2R$, at $R= 1/m\alpha = a_0$.  

    The Hamiltonian of the hyperfine interaction between the electron and proton is, 
\beq
  H_{hf} &=& \mu_e \,\vec\sigma_e \cdot \vec H(r),
\eeq 
where $\mu_e$ is the magnitude of the electron magnetic moment, $\vec\sigma_e$ is the electron Pauli spin matrix (twice the electron spin), and 
\beq
  \vec H(r) = -\vec\nabla \times (\vec\sigma_p \times \nabla)\frac{\mu_p}{r}
  \label{hproton}
\eeq
is the magnetic field produced by the proton magnetic moment, with $\vec\sigma_p$ the proton spin Pauli matrix.  Here $\mu_p= g_p e/4M_p$ is the magnitude of the proton magnetic moment,  with $g_p/2 \approx 2.79 $.  Then
\beq
  H_{hf} &=& -\mu_e \mu_p\left(\vec\sigma_e \cdot\vec\sigma_p \nabla^2  -  \vec\sigma_e\cdot\vec \nabla  \vec\sigma_p\cdot\vec \nabla \right)\frac1{r}. 
\eeq
In an s-state,  $\vec\sigma_e\cdot\vec \nabla  \vec\sigma_p\cdot\vec \nabla (1/r)$ averages to  $\frac13 \vec\sigma_e\cdot \vec\sigma_p \nabla (1/r)$, so that with $\nabla^2 (1/r) = -4\pi\delta(r)$ and the state~\eqref{phicoul}, one recovers the Fermi formula for the expectation of the hyperfine interaction.
\beq
   E_{hf} &= &\frac{8\pi}3  \mu_e \mu_p\big\langle  \vec\sigma_e \cdot \vec \sigma_p  \rangle |\varphi(0)|^2 
  =  \frac{8\mu_e \mu_p}{3 R^3}\big\langle  \vec\sigma_e\cdot \vec \sigma_p  \rangle.  
   \label{hf}
\eeq
In a spin singlet, $\langle\vec\sigma_e\cdot \vec \sigma_p  \rangle= -3$, and thus
 \beq
   E_{hf,0} &= & - \frac{8\mu_e \mu_p}{R^3}.
   \label{hf7}
\eeq
For $R = a_0$, this result is the usual hyperfine interaction. 
 
    With the hyperfine interaction added to the usual terms in Eq.~\eqref{E0R}, the expectation value of the total energy is
\beq
  E(R) = \frac{1}{2mR^2} - \frac{\alpha}{R} - \frac{8\mu_e \mu_p}{R^3},
  \label{naive}
\eeq
which has a local minimum in $R$ essentially at the Bohr radius, but diverges to arbitrarily negative values as
$R\to 0$.  

   The puzzle of whether the hyperfine interaction can destabilize the system to sizes far below the Bohr radius is intrinsically relativistic, since in non-relativistic limit $c\to\infty$, the electron magnetic moment $e\hbar/2mc$ itself vanishes.  Deriving the properties of bound relativistic Coulomb systems of two fermions, each of finite mass, can in principle be accomplished by  solving the relativistic 
 Bethe-Salpeter two body equation \cite{Bethe-Salpeter}.  Considerable effort has been devoted to reducing the problem of these systems to a description in terms of an equal time 4 $\times$ 4 matrix wave function in Dirac and real space. References~\cite{crater,Darewych2,Darewych,pilkuhn} representative a sample of work in this area.  To understand the flaw in the heuristic argument above, we analyze the relativistic atomic Dirac state using a variational wave function of spatial extent $R$, based on the Gordon solution \cite{Gordon} of the Dirac equation with an infinitely massive point proton, supplemented by the minimax principle \cite{talman}. 

   A central result is that, after the lower Dirac components are extremized at a specified radius, the fermion magnetic moment 
 is proportional to the inverse of the particle mass, e.g, $e/2m$ for the electron, only at the radius that minimizes the energy.  On the other hand, for all $R$ less than $1/m$ and arbitrary $Ze^2 < 1$l the Dirac energy has no minimum in $R$, and   
 the effective magnetic moments of point fermions in Coulombic atoms, including positronium and muonium (the $\mu^+e^-$ atom), become proportional to $R$ at small $R$, vanishing as $R\to 0$.  Thus the apparent $\sim -Z e^2/R^3$ divergence in the spin-singlet hyperfine interaction is softened to a $\sim -Z e^2/R$ divergence, which is dominated by the $+1/R$ divergence of the kinetic energy.  
 
   The magnetic moment of a point proton in hydrogen is also bounded for $R$ smaller than $1/M_p$ by  $g_p eR/4$, leading to a hyperfine energy softened to $\sim - g_p \alpha/2R$, smaller in magnitude than the electron kinetic energy.   In physical hydrogen though the finite size of the proton by itself removes the apparent divergence of the hyperfine intereraction as $R\to 0$,  as Miller \cite{miller},
responding to this paper, has shown in detailed calculations with the usual dipole form of the Sachs magnetic form factor of a finite proton.  Fnite size effects cannot however resolve the hyperfine puzzle in muonium or positronium, since charged leptons are point particles.  The relevant physics remains the Dirac suppression of the effective magnetic moments in trial configurations smaller than the Compton wavelengths of the participating fermions.  In neither hydrogen, nor positronium, nor muonium
does the hyperfine interaction lead to collapse.
 
We also look below (Sec.~\ref{Q}) at the hyperfine puzzle from the point of view of the potential felt by the electron at short distances.   One might be tempted to interpret the Fermi result, that the hyperfine energy is proportional to the square of the wave function at the origin, as saying that in the hyperfine interaction the electron feels a three dimensional delta function potential at the origin, $V_{hf,0} = -8\pi \mu_e\mu_f \delta(\vec r\,)$, for a spin-singlet.   However, one must treat such a potential, as with general contact interactions in low energy atomic interactions, only as an effective interaction or pseudopotential (see, e.g., Ref.~\cite{huang}) that cannot be used in higher order calculations, for example, of its bound states.\footnote{A real three-dimensional delta-function potential, realized as the limit of a spherically symmetric square well of radius $r_0 \to 0$ \cite{geltman}, has an arbitrarily large number of bound states of arbitrarily strong binding, whose existence would erroneously lead one to conclude that hydrogen would collapse.}   The actual attractive potential felt by an electron in a singlet state at short distances that arises from the hyperfine interaction cannot, as we show, produce additional bound states.

     In the next Section,~\ref{review}, we show how one can derive the exact energy 
   of the s-wave ground state Dirac wave function via the minimax variational procedure.   In Sec.~\ref{ps}, we apply the same variational method to the ground state of positronium.     Then in Sec.~\ref{magmoment} we derive the energy of an electron in an external magnetic field as well as in the magnetic field of a central, positively charged 
     spin-1/2 fermion, and hence find the effective (or ``shielded" \cite{hegstrom}) electron magnetic moment in terms of the variational wave function.   We briefly examine in Sec.~\ref{Q}, the actual equivalent potential felt by the electron from the hyperfine interaction.  
A key takeaway from this section is that the spin-independent contribution to the vector potential from the companion dipole moment is crucial for treating the small $r$ behavior of the Dirac wave function.  
We summarize and conclude in Sec.~\ref{summary}.  Appendix~\ref{solving} reviews the derivation of the exact Coulomb wave function for hydrogen ground state, and in Appendix~\ref{free} we illustrate the minimax principle for a free Dirac particle.  Finally in Appendix~\ref{unequal} we study the ground state energy, and the approximate wave function for two oppositely charged fermions of general masses with a Coulomb interaction.

    The framework we develop in this paper is useful for the problem of constructing diquarks in hadrons and in low density quark matter. There the QCD color electric interaction plays the role of the Coulomb potential and the QCD color magnetic interaction the role of the hyperfine interaction, leading to the same problem of the $R$ dependence of the hyperfine interaction as in hydrogenic atoms.  The diquark problem requires understanding the regime in which the strong interaction fine structure constant, $\alpha_s$, can be of order unity; thus the technology we develop to resolve the behavior of the hyperfine interaction in hydrogenic atoms becomes directly applicable in improved analytic treatments of diquarks \cite{diquark}.

  \section{Variational approach to the Gordon solution}
  \label{review}

    To uncover the roots of the hyperfine puzzle we work in terms of the relativistic Gordon solution of the Dirac equation for the Coulomb problem \cite{Gordon,Bethe-Salpeter}, and its extension to positronium, employing a simple variational method based on the exact hydrogenic wave function.   The Hamiltonian of the relativistic spin-1/2 Coulomb problem is, 
\beq
  H_H =  \gamma^0 \vec\gamma\cdot (\vec p+e\vec A(r)) + \gamma^0 m - \frac{\alpha}{r}
  \label{hcoul}
\eeq 
(the subscript $H$ is for hydrogen), 
where $r$ is the relative coordinate of the electron and Coulomb center, and $\vec A(r)$ the vector potential for 
the magnetic field felt by the electron (here $e>0$).   We include the hyperfine interaction subsequently.

  The exact  normalized Dirac ground state for an s-wave electron in the absence of a magnetic field is $\Psi (\vec r\,)  = (\phi(r),\chi(\vec r))$, with
\beq 
\phi = \frac{1}{\sqrt{1+X^2}}
\zeta
\tilde\varphi(r),\quad \chi=  iX\vec\sigma\cdot \hat r \phi.
\label{gndstate}
\eeq
Here $\zeta$ is the two component spinor in the upper Dirac components , e.g., $\zeta = \begin{pmatrix} 1\\0 \end{pmatrix}$ for a spin up electron, and
\beq
\tilde\varphi(r) = {\cal N} r^{\gamma-1}e^{-r/a_0},
\eeq
where $\gamma = \sqrt{1-\alpha^2}$ and  $\cal N$ is a normalization constant.  The derivation of the Dirac ground state energy and parameters is reviewed in Appendix~\ref{solving}. The $r^{\gamma -1}$ factor leads to a weak divergence at the origin~\cite{Bethe-Salpeter}, which generates, as Breit has shown \cite{breit}, higher-order \  corrections to the hyperfine interaction in the Dirac hydrogen atom.   The parameter $X$ is \cite{Bethe-Salpeter},
\beq
X= \frac{1}{\alpha}(1-\sqrt{1-\alpha^2}) \, >0, 
\label{X} 
\eeq
which is $\approx \alpha/2 $ for $\alpha \ll 1$.

  To discuss the system as a function of it size, we compute its energy   in the following using a variational trial wave function of the same form as \eqref{gndstate}, with
\beq
\tilde\varphi(r) = {\cal N} r^{\gamma-1}e^{-r/R},
\label{variational}
\eeq
and $X$, $\gamma$ and $R$ now parameters to be determined. 

   Variational approaches to solving the Dirac equation are beset by the ``variational catastrophe":  owing to the lack of a lower bound on the Dirac spectrum, simple extremization does not find a minimum energy, since one can always lower the energy indefinitely by mixing in negative energy states.  We adopt here the ``minimax"  procedure~\cite{talman} for the physical energy, where one first extremizes the ground state energy with respect to $X$ (the relative normalization parameter of the lower component) and then determines the radius by minimizing the energy with respect to $R$.  Extremization with respect to $X$  in fact leads to a maximum in the energy, a situation simply illustrated by the free Dirac particle, calculated variationally in Appendix~\ref{free}.   Once we have constructed a physical Dirac state, with energy $E(R)$, we determine the radius by minimizing the energy with respect to $R$.    The minimax  procedure is discussed in detail  in Ref.~\cite{talman} in terms of the positive and negative eigenvalues in the Dirac equation;  see also Refs.~\cite{kalf,kolakowska}, as well as Ref.~\cite{wu}
and references therein for alternate variational treatments of the Dirac hydrogen atom for general states.   Although the minimax method does not in general necessarily provide an exact upper bound to the energy,  calculations reported in the literature, e.g., \cite{talman,Darewych,Darewych2},  approximate the energy reasonably accurately.

   In the minimax procedure, $X$, which fixes the relative weight of the lower Dirac components, is first extremized at fixed $R$ and 
 $\gamma$.  The size $R$ is varied only after this minimax step.   To calculate the Coulomb energy in the variational state~\eqref{variational} we use,
\beq
\Big\langle \frac1{r}\Big\rangle = \frac{\int dr r^{2\gamma-1}e^{-2r/R}}{\int dr r^{2\gamma}e^{-2r/R}} = \frac1{\gamma R}.
  \label{1overr}
\eeq
The mass term gives,
\beq
  \langle \gamma^0 m\rangle = m\frac{1-X^2}{1+X^2}.
\eeq  
The expectation value of the kinetic energy -- since $\gamma^0\vec \gamma$ is simply the  $4\times 4 $ Dirac matrix
$
 \begin{pmatrix}
   0 & \vec \sigma  \\
   \vec\sigma & 0 \\
 \end{pmatrix},
$
with $\vec \sigma$ the Pauli matrices -- contributes,
\beq
  \langle \gamma^0 \vec\gamma\cdot \vec p\, \rangle &=& \frac{iX}{1+X^2}\int d^3r \tilde\phi \zeta^\dagger [\vec\sigma\cdot \vec p,\vec\sigma\cdot \hat r] \zeta\tilde\phi \nonumber\\&=& -\frac{2X}{1+X^2}\int d^3r \tilde\phi \,\frac{\partial\tilde\phi}{\partial r} = \frac{2X}{1+X^2}\frac1{R\gamma},  \nonumber\\
  \label{ke}
\eeq
where we use 
 \beq
  \frac{\partial}{\partial r} \tilde\phi =  \left(-\frac1{R} + \frac{\gamma -1}{r}\right)\tilde\phi
\eeq   
with Eq.~\eqref{1overr}.  Altogether the expectation value of the energy is,
\beq
 E_H=  \left( \frac{2X}{1+X^2} - \alpha\right) \frac{1}{R\gamma} + m \frac{1-X^2}{1+X^2}.
  \label{ecoulexact1}
\eeq

  Minimizing $E_H$ with respect to $X$ and fixed $R\gamma$,  the first step in the minimax procedure, gives
\beq
X^2+2XmR\gamma-1=0, 
\label{Xe}
\eeq
and thus
\beq
  X = -mR\gamma + \sqrt{(mR\gamma)^2 +1}.  
   \label{XR1}
\eeq
At this extremum in $X$, the energy depends only on the product $R\gamma$:
\beq
  E_H(R\gamma)= \frac1{R\gamma}\left(\sqrt{(mR\gamma)^2 +1} - \alpha\right).
  \label{Ec27}
\eeq
We next minimize $E_H$ as a function of $R$.     Since $dE_H/dX = 0$ at the extremum, the implicit $R\gamma$ dependence in $X$ drops out of the minimization.     Minimizing with respect to the explicit $R\gamma$ implies $1+X^2-2X/\alpha=0$, from which Eq.~\eqref{X} follows.   In addition, comparison of  Eqs.~\eqref{XR1} and \eqref{X} gives
 \beq
   m\alpha R\gamma = \sqrt{1-\alpha^2},
   \label{mrg7}
 \eeq
 and yields the form for the ground state energy,
\beq
E_H = m\sqrt{1-\alpha^2},
\label{ehma}
\eeq
which is, in fact, the exact solution obtained by solving the Dirac equation.

  The minmax procedure determines only $R\gamma$ and not $R$ and $\gamma$ separately. To obtain those separately we can match coefficients of $1/r$ in the Dirac equation (see Appendix~\ref{solving}) to find 
 \beq
   \gamma = \sqrt{1-\alpha^2},
 \eeq
and then $R = 1/m\alpha$, the Bohr radius.\footnote{An alternative procedure to matching coefficients in the Dirac equation is to make the ansatz -- useful for positronium as well -- that in the regime $R>1/m$, 
$\gamma = \sqrt{1-1/(mR)^2}$, so that $R\gamma= \sqrt{R^2-1/m^2}$.}
Note that 
\beq
E_H^2 = m^2 - 1/R^2,
\label{EmR}
\eeq
a result obeyed by the s-wave ground state of the Klein-Gordon hydrogen atom as well. 

    With the variational wave function  \eqref{variational} we can maximize the energy with respect to $X$ for given $R$.  However, the resulting energy at the maximum in $X$ has a minimum in $R$, for general  $Z\alpha$,  only if $Z\alpha<1$ and then $R>1/m$.  This latter condition is readily apparent from Eq.~\eqref{EmR}, where the energy becomes imaginary for $R<1/m$.  The limiting case $R=1/m$ occurs for $Z\alpha=1$ where $\gamma=0$.  As we see from Eq.~\eqref{Ec27},
\beq
\frac{dE_H}{dR} =  - \frac1{\gamma}\left(\frac1 {\sqrt{(mR\gamma)^2 +1}} -  Z\alpha\right);  
\eeq
since the first term in parenthesis is always $\le 1$, the derivative can vanish only if $Z\alpha \le 1$.   For  $Z\alpha <1$ and $\gamma = \sqrt{1-(Z\alpha)^2}$,
the derivative vanishes when $mRZ\alpha = 1$, as expected.  As $Z\alpha \to 1$, one has $\gamma \to 0$, at which point the wave function becomes singular at the origin.  For $Z \alpha>1$ there is never a minimum.   For  $ Z\alpha<1$,  one has $R>1/m$;  the variational state \eqref{variational} with $R<1/m$ cannot minimize the energy.

\section{Positronium}
\label{ps}

  We turn next to the ground state of positronium, whose Hamiltonian in the absence of the hyperfine interaction 
 is,\footnote{At this stage we omit transverse-photon terms in identifying how the variational state controls the effective magnetic moments.}
\beq
  H_{ps} &=&  \gamma^0_1 \vec\gamma_1\cdot \vec p_1 + (\gamma^0_1+\gamma_2^0) m+\gamma^0_2 \vec\gamma_2\cdot \vec p_2 - \frac{\alpha}{|\vec r_1- \vec r_2|},  \nonumber\\
  \label{hpos}
\eeq
where 1 denotes the electron and 2 the positron, and $m$ is again the electron mass.  
We write the overall ground-state trial wave function as a generalization of Eq.~\eqref{gndstate} for hydrogen, 
\beq
   \Psi_{ps} (r) & = &\Big(\frac{1}{\sqrt{1+X_1^2}}\left(\zeta_1, iX_1\vec\sigma\cdot \hat r \zeta_1\right) \nonumber\\&&
                                          \otimes  \frac{1}{\sqrt{1+X_2^2}}\left(\zeta_2,-iX_2\vec\sigma\cdot \hat r \zeta_2\right)\Big)
                                            \tilde\varphi(r), 
 \label{psgndstate}
\eeq
where $\zeta_1$ and $\zeta_2$ are the two component spinors of particles 1 and 2,
 $\vec r = \vec r_1- \vec r_2$, and $\tilde\varphi(r) \sim r^{\gamma-1} e^{-r/R}$.     While the components of the exact Coulomb wave function do not share a common spatial wave function ~\cite{Darewych}, the trial spatial wave function nonetheless remains a useful starting point.
  
   By symmetry $X_1=X_2 \equiv X$.  
Then the variational ground state energy is
\beq
 E_{ps}(\alpha) &=&   \langle \Psi_{ps}|H_{ps}|\Psi_{ps}\rangle\nonumber\\&& = \frac{2X}{1+X^2}\frac2{R\gamma} + 2 m \frac{1-X^2}{1+X^2} -  \frac{\alpha}{R\gamma}. 
    \label{psenergy00}
 \eeq
 Comparing with Eq.~\eqref{ecoulexact1} we see immediately that for this variational wave function,
\beq
   E_{ps}(\alpha) =  2E_H(\alpha/2)= 2m\sqrt{1-\alpha^2/4}.
 \label{epos}
\eeq
Numerical calculations of the ground state energy of the Hamiltonian \eqref{hpos}, for $0\le \alpha \le 2$,  give results extremely close to and straddling Eq.~\eqref{epos} for all $\alpha$ in this range \cite{Darewych,Darewych2}.

 Extremizing \eqref{psenergy00} with respect to $X$ again yields Eq.~\eqref{Xe}, from which it follows that at the minimum in $R$,
\beq
X &=& \frac{2}{\alpha}(1-\sqrt{1-\alpha^2/4}),
\label{xalpha} 
\eeq
the hydrogenic result with $\alpha$ replaced by $\alpha/2$.   In addition, $R= 2/m\alpha$ and $\gamma = \sqrt{1-\alpha^2/4}$.

     In the limit of small $\alpha$, in which the system is non-relativisic,  this equation yields the expected result,
\beq
   E_{ps}(\alpha) \to   2m\left(1-\frac18\alpha^2\right) = 2m - \frac12 m_r \alpha^2,
\eeq
where $2m$ is the total and $m_r = m/2$ is the reduced mass.

 \section{Magnetic moment of a particle in a Coulomb potential}
\label{magmoment}

  We turn now to the effects of a magnetic field on an electron in a Coulombic ground state, evaluating the expectation value of the term $\gamma^0 \vec\gamma\cdot e\vec A(r)$ in the Hamiltonian \eqref{hcoul}, and the corresponding term in positronium.     
 Similar to the calculation leading to Eq.~\eqref{ke},   we have,
\beq
 && \langle e \gamma^0\vec \gamma\cdot \vec A(r)\rangle
  = 2e\Re \int d^3r \phi^\dagger  \vec \sigma\cdot \vec A(r) \chi  \nonumber\\&&
 = \frac{eX}{1+X^2}  \int d^3r\phi^\dagger [ \vec \sigma\cdot \vec A(r), i\vec\sigma\cdot \hat r ]\phi \nonumber\\&&  =  \frac{2eX}{1+X^2} \int d^3r |\tilde\phi|^2  \hat r \times \vec A(r)\cdot\langle \vec\sigma \,\rangle.
\eeq

 The detailed response of the electron depends on whether $\vec A$ arises from a current source $\vec J(r)$ outside the atom or within the atom.
For a uniform external magnetic field, described by the vector potential,
$ \vec A(r) = \frac12(\vec H \times \vec r)$,
\beq
  \hat r \times \vec A(r) = \frac12 \hat r \times(\vec H \times \vec r) = \frac12 (\vec H - (\hat r\cdot\vec H) \hat r) r.
\eeq
Averaged over angles in an s-state, the right side becomes $\vec H  r/3$,
and with the wave function\footnote{More generally, for a one electron atom in its s-wave ground state in a spatially dependent external magnetic field,  the $H$ in Eq.~\eqref{constH} becomes simply the value of the field at the center of the atom.}  $\tilde \varphi(r)$,
\beq
 && \langle e \gamma^0\vec \gamma\cdot \vec A(r)\rangle
  =  \frac{eXR{\cal B}_o}{1+X^2}\,\langle\vec\sigma\, \rangle \cdot \vec H, 
  \label{constH}
\eeq
where
\beq 
   &&{\cal B}_o = \frac2{3R}\frac{\int d^3r  r \tilde \varphi^2}{\int d^3r  \tilde \varphi^2}= \frac{2\gamma+1}{3} \approx{1-\alpha^2/3},
   \label{bo}
\eeq
is a Breit correction to the electron $g$ factor.   As Eqs, \eqref{constH} and \eqref{bo} show, the energy shift due to a magnetic field depends separately on $\gamma$ and $R$,  unlike in the variational result \eqref{Ec27} for the ground state energy.

   The crucial takeaway of the above result is that the effective electron magnetic moment in a hydrogenic state, at the extremum in $X$, and prior to the minimization in $R$, is
\beq
      \mu_e^{\rm eff} = \frac{eXR}{1+X^2} {\cal B}_o.
 \label{effmagmom0}
\eeq
As we discuss below, the explicit factor $R$ softens the divergence of the hyperfine energy as $R\to 0$. 
The factor ${\cal B}_o$, which differs from unity because of the higher order corrections to $\gamma=1$ appearing in $\tilde \phi$, gives higher order corrections in $\alpha$; in the following we 
replace ${\cal B}_o$ with unity.{\footnote{While computation of terms of higher order in $\alpha$ is moot here, since the Dirac equation alone does not include radiative corrections, we note these effects in order to understand better hyperfine, or color magnetic, interactions in a relativistic treatment of diquarks \cite{diquark}.}}  
 
 With the energy $E_H$ or $E_{ps}$ extremized with respect to $X$,  Eq.~\eqref{Xe} implies
\beq
   \frac{X}{1+X^2} = \frac1{2\sqrt{(mR\gamma)^2+1}}.
   \label{xmrgamma}
\eeq
Thus for general $R$,
\beq
      \mu_e^{\rm eff}= \frac{eR}{2\sqrt{(mR\gamma)^2+1}}.
 \label{effmagmom7}
\eeq
As $R\to 0$, $\mu_e^{\rm eff} \to eR/2$, while at the minimum of the energy \eqref{Ec27}, which exists for  $R>1/m$.
$\mu_e^{\rm eff} = e/2m$, as expected.\footnote{The suppression of the magnetic moment at small radius does not rely on  the Coulomb factor $r^{\gamma-1}$.  With $\gamma\equiv 1$, the minimax equation becomes $X^2+2mRX-1=0$, and the extremized magnetic moment becomes $\mu_e^{\rm eff}=eR/\left(2\sqrt{(mR)^2+1}\right)$;  this magnetic moment goes continuously as a function of $mR$ from the correct value $e/2m$ at $R$ infinite to zero as $R$ goes to zero, an unphysical behavior. 
While the minimized energy, $m\sqrt{1-\alpha^2}$, is independent of $\gamma$, the  size without the $\gamma-1$ factor is smaller by a factor $\sqrt{1-\alpha^2}$ than the expected $1/m\alpha$.}
    
   We next consider the response to a magnetic field generated by currents at the center of the atom, as in the hyperfine interaction.  The vector potential produced by a point-like positively charged fermion (e.g., $f$ = p, $\mu^+$, or e$^+$) is, cf. Eq.~\eqref{hproton},  
 \beq
   \vec A(r) = \mu_f\vec\sigma_f \times \frac{\hat  r}{r^2}, 
   \label{Aptp}
\eeq
where $\mu_f$ is the positive fermion magnetic moment, $\vec\sigma_f/2$ is the fermion spin, and we neglect corrections arising from the motion of the positively charged fermion which even for positronium are only of order $\alpha$.
 In an s-state $\hat r \times \vec A(r)$ averages to $2\mu_f\vec\sigma/3r^2$, so that in the state $\tilde\varphi$ the hyperfine interaction becomes
\beq
 E_{hf} &=& -\langle e \gamma^0\vec \gamma\cdot \vec A(r)\rangle
    = - \frac{8 eXR{\cal B}_i}{3(1+X^2)} \frac{\mu_f}{R^3}\langle\vec\sigma_e\cdot\vec\sigma_f\rangle , \nonumber\\
 \label{hfgeneral}
\eeq  
where the Breit hyperfine correction here is
\beq
    {\cal B}_i =\frac{R^2}2 \frac{\int dr \tilde\phi^2}{\int dr r^2\tilde\phi^2} = \frac{1}{ \gamma(2\gamma-1)} \approx 1+3\alpha^2/2.
    \label{bi}
\eeq
We see in Eq.~\eqref{hfgeneral} the emergence of the explicit $1/R^3$ in the hyperfine energy, discussed in the Introduction.  Note that the Breit correction converges only if $\gamma>1/2$,

  In the following we neglect the Breit factors, and  consider at first the hyperfine interaction in positronium.  Then $\mu_{e^+} = -\mu_e$, and with Eq.~\eqref{hfgeneral},  
\beq
 E_{hf,ps} &=&
     \frac{8 X^2\alpha}{3(1+X^2)^2R}\langle\vec\sigma_e\cdot\sigma_{e^+}\rangle. 
\eeq  
Since $X/(1+X^2)$ is bounded by 1/2 in magnitude,  the hyperfine energy at small $R$ is in fact $\sim 1/R$, rather than the naive $1/R^3$.  With Eq.~\eqref{xmrgamma}, 
\beq
 E_{hf,ps} = \frac{2 \alpha}{3((mR\gamma)^2+1)R}\langle\vec\sigma_e\cdot\sigma_{e^+}\rangle,
\eeq
which is bounded in magnitude by the kinetic energy.
In the positronium ground state for small $\alpha$,  $R\to 2/m\alpha$,  and
\beq
   E_{hf,ps} &\to&  \frac{m\alpha^4}{12}  \langle  \vec\sigma_{e^+}\cdot \vec \sigma_{e^-}  \rangle, 
   \label{ehfps}
\eeq
in agreement with Refs.~\cite{KK,Bethe-Salpeter}; in a singlet state, $E_{hf,ps} \to -m\alpha^4/4$.

In muonium,  where the positive particle is pointlike, the trial state  \eqref{factorgndstate} for unequal masses implies that 
the response of the muon to an external magnetic field is described by the effective magnetic moment (cf. Eq.~\eqref{effmagmom0}), 
\beq
  \mu_\mu^{\rm eff}(R)   = -\frac{eX_\mu R}{1+X_\mu^2}.
 \label{peffmagmom0}
\eeq
Here $X_\mu$ measures the size of the lower Dirac components of the muon.  Since $X_\mu/(1+X_\mu^2) \le 1/2$, with $X_\mu\to 1$ as $R\to 0$, the effective muon magnetic moment in the small $R$ limit becomes, in magnitude, $eR/2$.
With this shielding of $  \mu_\mu^{\rm eff}(R)$ in the non-equilibrium state at small $R$, the muonium hyperfine interaction scales  as $\pm \alpha/R$, and is always well bounded by the electron kinetic energy.  

   For hydrogen with a point proton, Eq.~\eqref{hfgeneral} without the Breit correction implies that in the singlet state of hydrogen -- with proton magnetic moment, $g_p e/4M_p$ -- the hyperfine energy is,
\beq
 E_{hf} &\approx& - \frac{\alpha g_p}{M_P R^2\sqrt{(mR\gamma)^2+1 }}.
    \label{ehfbreit1}
\eeq
For small $R$ this remains smaller in magnitude than the electron kinetic energy, $\sim1/R$, down to $R\approx \alpha g_p/M_p \approx$ 0.053 fm, well within the proton itself.   However, as mentioned in the Introduction, in physical hydrogen the finite size of the proton removes the apparent divergence of the hyperfine interaction as $R\to 0$ \cite{miller}. 

   Since the hyperfine interaction, for given $R$, scales with $\alpha$, we ask -- with a view to diquarks held together by color electromagnetic forces -- what is the maximum value of $\alpha$ for which a positronium-like system is 
 stable as $R\to 0$.  The total energy of positronium in its spatial ground state, including the hyperfine interaction, behaves there as
\beq
 E&\to& \frac1{R\gamma}\Big(2  -\alpha
 +\frac{2\alpha}{3} \frac{ \langle  \vec\sigma_{e^+}\cdot \vec \sigma_{e^-}  \rangle }{(2\gamma-1)}  \Big). 
 \label{ER0}
\eeq 
Stability as $R\to 0$ requires,
\beq
  2   > \alpha\left(1 -\frac{2}{3(2\gamma-1)}   \langle  \vec\sigma_{e^+}\cdot \vec \sigma_{e^-} \rangle \right) .
\eeq
The singlet state (where $\langle \vec\sigma_1\cdot \vec \sigma_2  \rangle = -3$) has a stronger negative contribution than the triplet state, and so the criterion for absolute stability, in a variational state specified by a $\gamma$, is that $\alpha$  be bounded above by $2(2\gamma-1)/(2\gamma+1)$.
 
\section{Spatial structure of the hyperfine potential}
\label{Q}

  As mentioned in the Introduction, the hyperfine interaction \eqref{hf} can be understood as the expectation value of a pseudopotential which is a delta function at the origin.  In this Section we  examine the conclusions of the previous sections from a related perspective -- by constructing the effective potential in the Dirac equation felt by the electron in the presence of the hyperfine interaction, with a Coulomb attraction to a massive central point fermion.  We follow Ref.~\cite{blinder} in spirit. 

   The full Dirac equation for the Coulomb potential with a vector potential, written out for the upper and lower components, is
 \beq
    \left(E - m +\frac{\alpha}{r}\right)\phi = \vec\sigma\cdot(-i\vec\nabla + e\vec A(r))\chi, \nonumber\\
    \left(E + m +\frac{\alpha}{r}\right)\chi = \vec\sigma\cdot(-i\vec\nabla + e\vec A(r))\phi.
    \label{diracA}
 \eeq
Eliminating $\chi$ we have
 \beq
 && \hspace{-12pt}(E +m)\left(E - m +\frac{\alpha}{r}\right)\phi \nonumber\\&&\hspace{-12pt} = \vec\sigma\cdot(-i\vec\nabla +e \vec A(r)){\cal Q}(r)\vec\sigma\cdot(-i\vec\nabla  +e\vec A(r))\phi, 
  \label{d-tem}
 \eeq
 where
 \beq
   {\cal Q}(r) \equiv \frac{E +m}{E  + m +\alpha/r}.
 \eeq
 
   The terms on the right proportional to the electron spin, with $(\vec\nabla \times \vec A(r)+ \vec A(r)\times \vec\nabla)\phi = \vec H(r)\phi$, where $\vec H(r)$ is the magnetic field,  become,
 \beq
       e\vec\sigma\cdot(\vec\nabla   {\cal Q}(r)\times \vec  A(r))\phi +e  {\cal Q}(r)\vec\sigma\cdot  \vec H(r)\phi,
       \label{twomagterms}
 \eeq
and with Eq.~\eqref{Aptp} for the vector potential of the positive central fermion, 
\beq
e\vec\sigma\cdot  \vec\nabla   {\cal Q}(r)\times \vec  A(r) = e \frac{\mu_f}{r^2}\frac{d  {\cal Q}}{dr} \vec\sigma\cdot\left(\vec\sigma_p   -(\vec\sigma_p \cdot \hat  r)\hat r \right).
\eeq
In a singlet state, to which we first confine our attention, $\sigma_{ei}\sigma_{pj} \to - \delta_{ij}$, and $\vec\sigma\cdot \vec H(r) = -8\pi \mu_f \delta(\vec r\,)$.  Since ${\cal Q}(0)=0$, the second term in~\eqref{twomagterms} vanishes, while
\beq
e\vec\sigma\cdot  \vec\nabla   {\cal Q}(r)\times \vec  A(r) \to -2e \frac{\mu_f}{r^2}\frac{d  {\cal Q}}{dr}.
\eeq
 Altogether,
 \beq
 && \hspace{-12pt}\left(E  - m +\frac{\alpha}{r}\right)\phi =\nonumber\\&&\hspace{-12pt}(-i\vec\nabla +e \vec A(r))\frac{{\cal Q}}{E +m} \cdot(-i\vec\nabla  +e\vec A(r))\phi  + V_{hf,0}(r)\phi, 
\nonumber\\
  \label{d-tem3}
 \eeq
 where 
 \beq
 \label{Vhf0}
  V_{hf,0}(r) &=&  -\frac{2e\mu_f }{E +m} \frac1{r^2}\frac{d  {\cal Q}}{dr} 
= -  \frac{2e\mu_f  }{r^2}   \frac{\alpha}{((E  + m)r +\alpha)^2}\nonumber\\
\eeq
is the effective potential an electron in a singlet state feels in its hyperfine interaction with the positive fermion.    

How is this very attractive $V_{hf,0}$ potential for a positive fermion $f$, $\sim -2\mu_f/er^2$ at small $r$, related to the apparent hyperfine puzzle mentioned in the Introduction?  In particular, can  $V_{hf,0}$ lead to  highly localized bound states near the origin, on the scale of the classical radius of the electron, or even an instability in the system?  Indeed, while the highly attractive potential might be construed as another manifestation of the hyperfine puzzle, it does not, as we show now, give rise to such localized bound states, thus confirming that the hyperfine puzzle is only a mirage. 

As Landau and Lifshitz discuss \cite{LLqm}, a $r^{-2}$ potential has no bound states if the coefficient $2\mu_f/e$ is less than $1/8m$, or $\mu_f < \mu_e/8$, clearly satisfied in hydrogen and muonium. 
For positronium, where such a $-1/r^2$ potential by itself would lead to total collapse of the system (independent of electron-positron annihilation), we must take into account the other terms in Eq.~\eqref{d-tem}.  Noting that $\hat r\cdot \vec A = 0$, we rewrite Eq.~\eqref{d-tem} as
  \beq
 && \hspace{-48pt}\left(\left(E   +\frac{\alpha}{r}\right)^2- m^2 +\vec\nabla^2 -\alpha \vec A(r)^2\right)\phi
  \nonumber\\&&\hspace{-12pt}  
 = -\frac{\alpha}{(E +m)r+\alpha}\left(\frac{1}{r}\frac{d\phi}{dr}
   +\frac{2e\mu_f}{r^3}\phi\right).
  \label{d-tem6}
 \eeq 
The repulsive spin-independent $\alpha\vec A(r)^2 = -\alpha \mu_f^2/r^4$  term dominates the hyperfine term, $-2e\mu_f/r^3$ on the right at distances $\lesssim \alpha/m_f$,preventing the formation of localized bound states.  In positronium the $A^2$ term wins at distances smaller than the classical radius of the electron.  See Ref.~\cite{himpsel} for more detailed analysis.   In all two fermion atoms, the hyperfine interaction, although quite attractive at small $r$, leads neither to localized bound states nor collapse.   
 
  To connect with the analysis of the hyperfine interaction in terms of the Dirac structure of the electron in the previous section, we note that for general spin the expectation value of the hyperfine interaction in an s-state is,
\beq
  E_{hf} =  \frac{8\pi}3 \langle \vec\sigma\cdot\vec \sigma_p\rangle \frac{e\mu_p }{E_H+m} 
 \int dr  \frac{d  {\cal Q}}{dr}|\phi(r)|^2.
  \label{bhf}
\eeq
This form allows $\phi(r)$ to have a weak divergence at the origin (a difficulty with the Fermi expression \eqref{hf}; see Ref.~\cite{Bethe-Salpeter}) and still give a finite energy.  
Integrating by parts on the right side of Eq.~\eqref{bhf} and using Eq.~\eqref{X} we find,
\beq
&&\frac1{E_H +m}\int dr \frac{d  {\cal Q}}{dr}|\phi(r)|^2\nonumber\\&&  \nonumber\\&&
=  \int dr\frac{2}{E_H  + m +\alpha/r} \left(m\alpha + \frac{1-\gamma }{r}\right) \phi(r)^2 \nonumber\\&&
= 2X \int dr  \phi(r)^2.
\eeq
Then with the normalization of $\phi$, we have
\beq
 && \hspace{-48pt} \int dr  \phi(r)^2 = \frac{\int dr  \phi(r)^2}{(1+X^2)\int 4\pi r^2 dr  \phi(r)^2} \nonumber\\&&
     =  \frac{1}{2\pi a_0^2(1+X^2)} \frac{1}{\gamma(2\gamma-1)},
  \eeq
and since $eXa_0/(1+X^2) = \mu_e$, we recover the hyperfine energy in the form in\footnote{To see the origin of the Fermi form of the hyperfine interaction, we note that $d  {\cal Q}/dr$ has a range of order the classical radius of the electron, $\alpha/m$, and in addition, $\int dr d  {\cal Q}/dr = 1$.
Thus $d{\cal Q}/dr \approx \delta^{(1)}(r)$, so that Eq.~\eqref{bhf} gives the hyperfine interaction as essentially proportional to the probability of finding the electron at the origin.} Eq.~\eqref{hf}.

\section{Summary and conclusion} 
\label{summary}

    We have addressed here the puzzle raised by the Fermi result for the hyperfine energy of hydrogen or positronium, that in an atom of size $R$, the hyperfine interaction scales as $\pm1/R^3$, raising the question of why in a spin singlet state (attractive sign) these systems do not collapse.  Studying the Dirac Coulomb problem using the minimax variational approach, we show:

 1) For radii $R>1/m$: The magnetic response of the electron, in a state that minimizes the energy with respect to radius, is governed by $\mu_e=e/2m$, with the Breit correction $1/\gamma(2\gamma-1)$ to the hyperfine energy, where the spatial wave function is $\sim r^{\gamma - 1} e^{-r/R} $.   By contrast, in an external magnetic field $H$, the electron $g$ factor receives a Breit correction $(2\gamma +1)/3$. 

  2) At small $R<1/m$:  There is no minimum in the variational energy as a function of $R$, and the {\it effective} electron magnetic moment scales as $R$ instead of $1/m$, softening the naive singular behavior.  With the proton magnetic moment treated within the Dirac framework, the hyperfine interaction behaves as $\pm1/R$, for $R<1 /M_p$;  the hyperfine interaction is not strong enough at small $R$ to induce a collapse of hydrogen.  
  
  3) The hyperfine interaction in positronium is similarly softened for configurations smaller than the electron Compton wavelength.
 
  In studying this problem we have introduced a variational approach to calculating the ground state of hydrogen, and extended it to the Coulomb groiund state for finite mass fermions, in the text for positronium, and in Appendix \ref{unequal} to arbitrary mass fermions.
 
    We have not given here explicit results for the energy and radius of positronium or hydrogen-like bound states that are accurate to higher order in 
 $\alpha$, since other physics, e.g., vacuum polarization, contributes beyond that captured by the solution to the Dirac potential problem.  However the dependence of the energy and radius on terms higher order in the fine structure constant become important in the framework of relativistic diquarks interacting via color electric and magnetic interactions, a problem we treat in a subsequent paper \cite{diquark}. 
   
\section*{Acknowledgments}
This research has been supported by the Simons Foundation and NSF Grant Nof013199 to author GRF, and carried out at the Aspen Center for Physics, supported by NSF Grant No. PHY-2210452.  Author GB is grateful to the late Geoff Ravenhall for sharing his wisdom about relativistic atoms, and to Christopher Pethick, Tetsuo Hatsuda and Wolfgang Ketterle for  insightful discussions.

\appendix

\section{Derivation of the hydrogen atom ground state solution of the Dirac equation}
\label{solving}

  In this Appendix we very briefly summarize how the Gordon s-wave ground state for a spin-up electron, Eq.~\eqref{gndstate}, with
$
 \tilde\varphi (r) = {\cal N}e^{-r/R}r^{\gamma -1}
$,
exactly solves the Dirac equation. The two usual coupled Dirac equations [\eqref{diracA} with $\vec A = 0$]  for the lower and upper components, $\phi$ and $\chi$,  reduce  for the forms in Eq.~\eqref{gndstate} to,
\beq
     \left(E_H- m + \frac{\alpha}{r}\right) \tilde\varphi &=& (-i\vec\sigma\cdot\vec\nabla) ( i\vec\sigma\cdot \hat r) \tilde\varphi   X \nonumber\\
    && \hspace{-48pt}   =X \left( - \frac1{R} +\frac{\gamma+1}{r}  \right)\tilde\varphi,
 \label{e-m}
 \eeq
since $\vec\nabla \cdot \hat r = 2/r$;
and in addition,
\beq
  \left(E_H+m + \frac{\alpha}{r}\right) X  =  \left(\frac1{R}+ \frac{1-\gamma}{r}\right).
  \label{e+m}
\eeq

  Matching term by term in $r$, we have from Eq.~\eqref{e+m},
\beq
X = \frac{1}{(E_H+m)R} =  \frac{1-\gamma}{\alpha}.
\label{xemr}
\eeq
while Eq.~\eqref{e-m} implies that
 \beq
   E_H- m =  -X/R,
 \label{exmr}
 \eeq
 and
 \beq
  X= \frac{\alpha}{\gamma+1}.
   \label{axg}
 \eeq
 
   Eliminating $X$ between Eqs.~\eqref{xemr} and \eqref{exmr}, we have
 \beq
 E_H^2 = m^2  -1/R^2,
 \eeq
 and eliminating $R$ as well,
 \beq
   E = m\gamma.
 \eeq
Then Eqs.~\eqref{xemr} and \eqref{exmr} imply
  \beq
   \gamma = \sqrt{1-\alpha^2},
 \eeq
 with  $E_H<m$ implying the positive square root.
 Finally, 
 \beq
   R = 1/m\alpha = a_0.
 \eeq

\section{Free Dirac particle}
\label{free}

\begin{figure}[h]
\includegraphics*[width=1.1\linewidth]{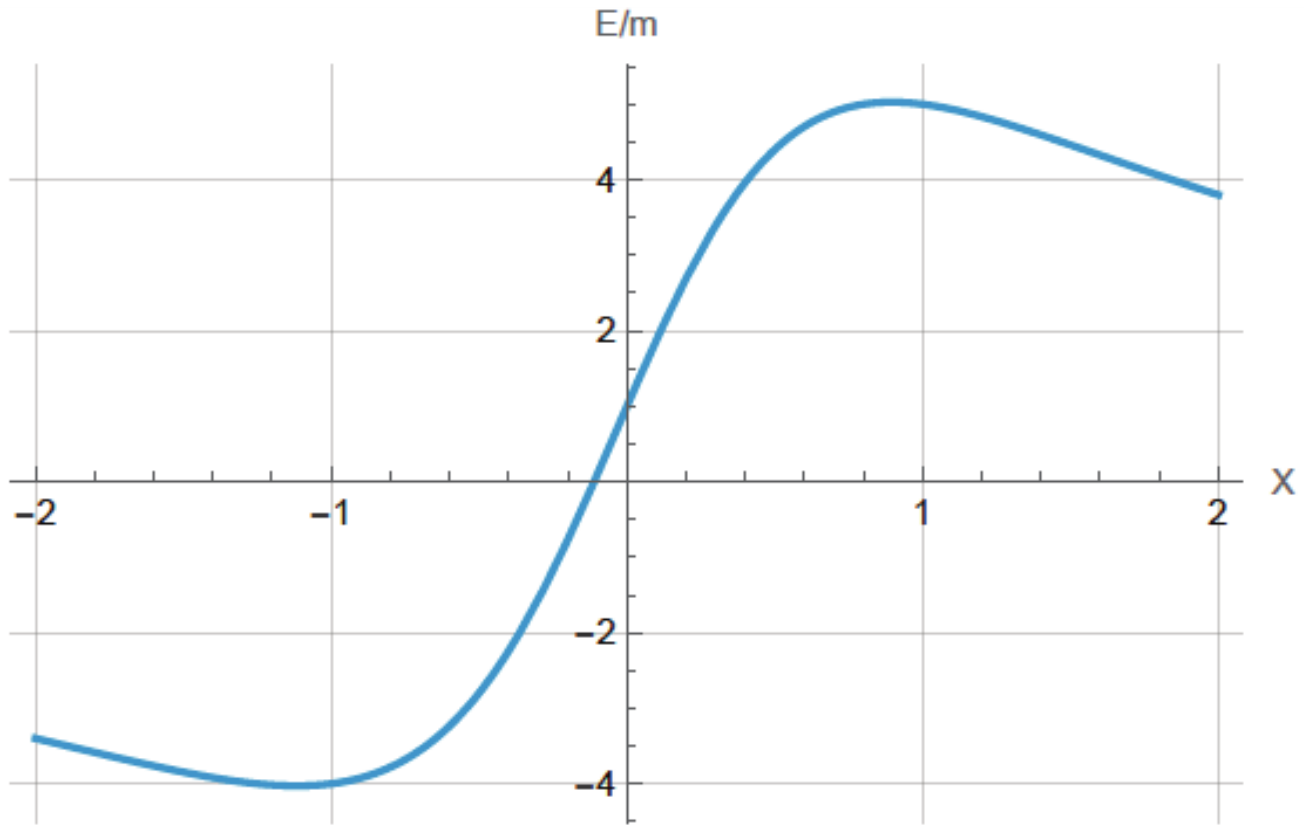}
\caption{$E/m$ vs.~$X$ for $p = 5m$.  The maximum at positive $X$ is the positive energy free particle solution, while the minimum at negative $X$ is the negative energy free particle solution.
}
\label{EvsX}
\end{figure}

 For a free Dirac particle of momentum $p$, with trial wave function $\Psi(r)$ given in \eqref{gndstate},  the expectation value of the Hamiltonian $H = \gamma^0(\vec\gamma\cdot \vec p +m)$ is,
\beq
  E =  \frac{2Xp+ m(1-X^2)} {1+X^2}. 
\eeq
Extremizing with respect to $X$ we find
\beq
 pX^2+ 2mX -p = 0.
\eeq
Thus
\beq
  X = \frac{E-m}{p}= \sqrt{\frac{E-m}{E+m}},
\eeq
where $E=\sqrt{p^2+m^2}$,  recovering the exact free particle Dirac spinor.

  Figure~\ref{EvsX}, a plot of $E$ vs. $X$, illustrates  the maximum at positive  $X$, corresponding to a positive energy free particle solution, and the minimum at negative $X$, corresponding to a negative energy free particle solution.

\section{Fermions of different masses, with Coulomb attraction}
\label{unequal}

The Dirac Coulomb problem for two fermions with finite masses, $m_1$ and $m_2$, with Hamiltonian,
 \beq
  H &=&  \gamma^0_1 \vec\gamma_1\cdot \vec p_1 + \gamma^0_1 m_1+\gamma^0_2 \vec\gamma_2\cdot \vec p_2 + \gamma^0_2 m_2 - \frac{\alpha}{r},\nonumber\\
\eeq
where $r= |\vec r_1-\vec r_2|$,  has not been solved analytically. The only exact solution is when one of the masses is infinite. 

 Nonetheless we can get important information about the energy of the ground state if we assume that for large $r$
the two particle $4\times4$ Dirac wave function is a product of individual wave functions of the form \eqref{gndstate},  
\beq
   \Psi & = &\Big\{\frac{1}{\sqrt{1+X_1^2}}\left(\zeta_1, iX_1\vec\sigma\cdot \hat r \zeta_1\right) \nonumber\\&&
                                          \otimes  \frac{1}{\sqrt{1+X_2^2}}\left(\zeta_2,-iX_2\vec\sigma\cdot \hat r \zeta_2\right)\Big\}
                                            \bar\varphi(r),
 \label{factorgndstate}
\eeq
where  $\zeta_1$ and $\zeta_1$ are the two component spinors of particles 1 and 2, times a common spatial wave function.   
We also assume that for $r$ much larger than the size, $R$, of the bound state, that  $d\bar\phi/dr =  -\bar\phi/R$ plus terms that fall off faster in $r$ and can thus be neglected asymptotically.

   In terms of $2\times2$ blocks, $\Psi$ can be written as,
 \beq
 \Psi_{11} &=& (\zeta_1\otimes  \zeta_2)  {\cal Z}\nonumber\\  
 \Psi_{21} &=& ( iX_1\vec\sigma\cdot \hat r \zeta_1\otimes  \zeta_2)  {\cal Z} \nonumber\\ 
 \Psi_{12} &=&  (\zeta_1\otimes -iX_2\vec\sigma\cdot \hat r \zeta_2 ) {\cal Z}   \nonumber\\ 
 \Psi_{22} &=&  (iX_1\vec\sigma\cdot \hat r\zeta_1\otimes  -iX_2\vec\sigma\cdot \hat r \zeta_2) {\cal Z},  
\eeq
where
\beq
  {\cal Z}(r) = \frac1{\sqrt{1+X_1^2}\sqrt{1+X_2^2}}  \bar\varphi(r). 
\eeq

   In the  two particle Dirac equation for large $r$, the Coulomb potential can be neglected, and the equation becomes
 \beq
  (h_1\otimes 1 + 1 \otimes h_2)\Psi = E\Psi,
  \label{2pd}
\eeq
where in the particle 1 subspace,
\beq 
  h_1  &=& \gamma^0_1 \vec\gamma_1\cdot \vec p_1 + \gamma^0_1 m_1 = \to 
\begin{pmatrix}
m_1&i\vec\sigma\cdot \hat r/R \\ i\vec\sigma\cdot \hat r/R &-m_1 \end{pmatrix}, \nonumber\\
\eeq
and similarly in the particle 2 subspace
\beq 
  h_2  &=& \gamma^0_2 \vec\gamma_2\cdot \vec p_2 + \gamma^0_2 m_2  \to 
\begin{pmatrix}
m_2&-i\vec\sigma\cdot \hat r/R \\ -i\vec\sigma\cdot \hat r/R &-m_2 \end{pmatrix}.\nonumber\\
\eeq

  Component by component, Eq.~\eqref{2pd} becomes the algebraic equations,
\beq
&& (-E + m_1 +m_2) -(X_1+X_2)/R  = 0, \nonumber\\
&&(-E  -m_1-m_2)X_1X_2 +( X_1+X_2)/R  =0, \nonumber\\
&&  ( -E+m_1-m_2)X_2 -(X_1X_2-1)/R = 0,\nonumber\\
&& (-E-m_1+m_2)X_1 -(X_1X_2-1)/R  = 0.
\eeq
The top two equations give
\beq
&&X_1X_2 -1 =  \frac{-2E} {m_1+m_2+E},
\eeq
and
\beq
 (X_1+X_2)/R  =  m_1 +m_2-E,
\eeq
while the bottom two equations give
\beq
    X_1+X_2 =  \frac{-2E}{ E^2-(m_1-m_2)^2} \frac{X_1X_2-1}{R}.
    \label  {X+X}
\eeq
Eliminating the $X$'s from these three equations we have
\beq
  E^4 -2(m_1^2+m_2^2- 2/R^2)E^2 + (m_1^2-m_2)^2,
\eeq
a quartic equation with the remarkably simple physical solution,
\beq
   E = \sqrt{m_1^2 -1/R^2}+\sqrt{m_2^2 -1/R^2},
   \label{magic}
\eeq
for arbitrary $m_1$ and $m_2$. 

We emphasize that the only assumptions going into this result are that the two particle state at large distances factorizes as \eqref{factorgndstate}, and that the overall spatial wave function has a common exponential behavior $\sim e^{-r/R}$ at large $r$.   The result \eqref{magic} can be derived as well from the more complete analysis of Ref.~\cite{Darewych}.   

For positronium, we recover $E^2 =  4(m^2- 1/R^2)$, and for hydrogen, where $m_2\to \infty$, 
\beq
 E-m_2 = \sqrt{m_1^2  -1/R^2},
 \label{magich}
\eeq
as expected.

\end{document}